\documentclass[sn-mathphys-num]{sn-jnl}

\usepackage{graphicx}%
\usepackage{multirow}%
\usepackage{amsmath,amssymb,amsfonts}%
\usepackage{amsthm}%
\usepackage{mathrsfs}%
\usepackage[title]{appendix}%
\usepackage{xcolor}%
\usepackage{textcomp}%
\usepackage{manyfoot}%
\usepackage{booktabs}%
\usepackage{algorithm}%
\usepackage{algorithmicx}%
\usepackage{algpseudocode}%
\usepackage{listings}%

\theoremstyle{thmstyleone}%
\theoremstyle{thmstyletwo}%

\theoremstyle{thmstylethree}%

\begin{document}

\title[Article Title]{Cylindrical gravitational impulse wave}


\author*[1,2]{\fnm{Jean J. Defo}}\email{defo.jeanjules@yahoo.fr}
\author[1,2]{\fnm{Victor K. Kuetche} }\email{vkuetche@yahoo.fr}
\affil*[1]{\orgdiv{National Advanced School of Engineering}, \orgname{University of Yaounde I}, \orgaddress{\postcode{P.O. Box 8390}, \country{Cameroon}}}
\affil[2]{\orgdiv{Department of Physics}, \orgname{Faculty of Science, University of Yaounde I}, \orgaddress{\postcode{P.O. Box 812}, \country{Cameroon}}}


\abstract{The description of gravitational waves as explosion and implosion waves as predicted by Weber and Wheeler [{\it Rev. Mod. Phys. {\bf 29} 509 (1957)}] in Einstein and Rosen spacetime, has recently been confirmed following observations by the LIGO-VIRGO scientific team [{\it Phys. Rev. Lett. {\bf 116 } 061102 (2016)}] resulting from the collision of two massive black holes. In this dynamics, we explore a new possibility in the construction of gravitational waves like explosion and implosion waves, the special case of Jordan and Ehlers spacetime, by studying the exact solutions of the Einstein field equations. For this purpose, we use the inverse scattering method of Pomeransky in association with the method of Piran et al. [{\it Phys. Rev. D {\bf 32} 3101 (1985)}] by solving the Einstein field equations in combination with the specific metric derived from Jordan and Ehlers in order to obtain a two-soliton solution with complex conjugate poles that we assimilate to the gravitational wave. Consequently, under certain conditions, we obtain the Einstein and Rosen waves and the Chandrasekhar transcendental waves.}


\maketitle

\subsection{Introduction}
The existence of gravitational waves in Einstein's theory of general relativity has led to renewed interest in understanding and interpreting their exact solutions to the field equations \cite{I1,I2,I3,I4}. Investigating the consequences of the theory of general relativity, Einstein came to regard gravitational waves as a wave solution to his equations in the weak-field approximation. A few years later, in their first attempt to construct the gravitational waves revealed by Kennefick \cite{I2}, Einstein and Rosen found a family of exact solutions to the field equations, which did not, however, satisfy the harmonic conditions and contained singularities. This statement was later contradicted in a paper published by Einstein and Rosen \cite{I3} concerning the veracity of these gravitational waves. Concerning these exact solutions, there are two of them: the monochromatic wave and the impulse wave \cite{I4}. We note that the existence of the impulse wave will be the subject of several investigations by several authors including: Rosen \cite{I5,I6}, Piran and al. \cite{I7} and Halilsoy \cite{I8}. We note that in the works contained in ref \cite{I4,I5} there is a discrepancy of interpretation concerning the existence of the gravitational energy during the propagation of the impulse wave. The information related to this discrepancy, results from the fact that the use of polar coordinates instead of Cartesian coordinates \cite{I6} makes the density of gravitational energy be totally null \cite{I4,I5}. This approach will be confirmed by a new numerical method of solving the exact solutions of the field equations \cite{I7}, which validates the different results obtained analytically \cite{I5,I8}. As we have indicated that the problem of the theory of gravitation is based on exact solutions of the field equations, highlighting the multitude of different methods of solving the problems posed by Einstein's general relativity, we can see that there are more powerful methods than others \cite{I9}.  Considering the inverse scattering method (ISM) of Belinski and Zahkarov \cite{I10}, the idea is actually to write the solutions of black holes and gravitational waves in terms of the addition of such solitons. In particular, the Kerr black hole is the addition of two solitons. Use this method \cite{I10} allowed to obtain two solutions, namely the monosoliton and two-soliton.  These two solutions were used with the metric of Jordan and Ehlers \cite{I11}, to highlight the phenomenon of Faraday analogue \cite{I12}. We note that it was also used in association with the metric of Einstein and Rosen \cite{I3} whose application allowed them to highlight the existence of cosmic chains \cite{I13} and the notion of time shift \cite{I14}. Numerous phenomena in the field of gravitation and cosmology with incredible results have been revealed by applying this method in its mathematical development. These phenomena are widely discussed within the framework of the one-soliton system and a two-soliton system with promising prospects \cite{I15}. The one-soliton system has shown to be a valuable tool in this study \cite{I15} for comprehending cosmic waves in the Friedmann model \cite{I16} and its implications for cosmology. Similarly, the two-soliton system in cosmology provides a further benefit since it allows us to provide a detailed description of the universe's initial inhomogeneity, which in this stage will partially permit the development of gravitational waves \cite{I17}. Although the ISM above \cite{I10} is interested in solving field equations, we note that the latter is used to solve problems in dimension two. A few years ago, this method was improved in the form of Pomeransky's ISM \cite{I18} in the construction of black holes. The central idea of Pomeransky \cite{I18} work is to simplify the construction of black hole solutions in higher dimensions via the inverse scattering method, and this is a different set with respect to the one the authors are using \cite{I10}. It was also used for the first time in the explanation of such phenomena as: Faraday analogue, time shift \cite{I19,I20}. In this study, we illustrate the behavior of gravitational waves as explosion and implosion waves in the specific spacetime of Jordan and Ehlers \cite{I11} using the two-solution system of cylindrical solitons discussed by Tomizawa and Mishima\cite{I19}, whose mathematical model is based on (ISM) Pomeransky's \cite{I18}. Using the decomposition method of Piran et al. \cite{I7}, we create a signal with qualities comparable to those acquired by the scientific team LIGO-VIGO \cite {I21}. The organisation of the paper is settled as fellows : in the next section \textbf{2}, we present the form of the specific Jordan and Ehlers metric \cite{I11} as well as the field equations governing the behavior of the momentum wave. Concerning section \textbf{3}, we present the gravitational soliton obtained by ISM of Pomeransky \cite{I18} which we assimilate to the gravitational impulse wave. In this section, we will study the propagation of the wave in the form of explosion and implosion according to the time coordinate $t$ and the radial coordinate $\rho$ as well as the different forms of gravitational energy density. In section \textbf{4} is devoted to conclusion.

\subsection{Basic Equations}\label{sec2}
Let us write down the general the Jordan and Ehlers \cite{I11} metric characterizing the interaction of two gravitational waves and the four field equations as follows:
\begin{equation}
ds^2  = e^{2(\gamma  - \psi )} (d\rho ^2  - dt^2 ) + \rho ^2 e^{ - 2\psi } d\phi ^2  + e^{2\psi } (dz + \omega d\phi )^2,
\end{equation}
 \begin{equation}
 \psi ,_{tt}  - \frac{{\psi ,_\rho  }}{\rho } - \psi ,_{\rho \rho }  = \frac{{e^{4\psi } }}{{2\rho ^2 }}(\omega ,_t^2  - \omega ,_\rho ^2 ),
 \end{equation}
 \begin{equation}
\omega ,_{tt}  + \frac{{\omega ,_\rho  }}{\rho } - \omega ,_{\rho \rho }  = 4(\omega ,_\rho  \psi ,_\rho   - \omega ,_t \psi ,_t ),
\end{equation}
 \begin{equation}
\
\gamma ,_\rho  = \rho(\psi ,_t ^2  + \psi ,_\rho ^2 ) + \frac{{e^{4\psi } }}{{4\rho}}(\omega ,_t ^2  + \omega ,_\rho ^2 ),
\
\end{equation}
 \begin{equation}
\
\gamma ,_t  = 2\rho\psi ,_t \psi ,_\rho  + \frac{{e^{4\psi } }}{{2\rho}}\omega ,_t \omega ,_\rho,
\
\end{equation}
where $(\rho ,z,\phi )$ represents the cylindrical coordinates and $t$ the time. The different arbitrary functions $\psi$, $\omega$ and $\gamma$ depend on $\rho$ and $t$. It is also noted that the previous quantities written with comma as subscript denote their partial derivatives with the associated variables. It is important to point out that the functions $\psi $, $\omega $, and $\gamma $ derived from the general metric and Einstein's field equations identified by equations (1)-(5) each have a specific role, particularly $\psi $ and $\omega $ represent the two dynamic degrees of freedom of the gravitational field in the theory of general relativity, so that $\psi $ corresponds to the $''+'' $ mode and $\omega $ to the $''\times'' $ mode. As for the $\gamma $ function, in these various equations mentioned, it plays the role of total gravitational energy per unit length between the axis of symmetry and the $\rho $ radius at time $t $, ref. \cite{I7}, whose behavior during the propagation of the gravitational wave through spacetime will be discussed in full.
\\
\\
In a recent study by Tomizawa and Takashi \cite{I20} developed for the search of new gravitational soliton waves, they pointed in the conclusion section of this paper one important query that needs to be clarified, i.e. the phenomenon  where the $''+'' $ mode waves rapidly decreases owing to the nonlinear effects while the cross $''\times''$ mode behave arbitrary. This query actually constitutes our motivation in this work. As far as we are concerned, we strongly believe  that in order to provide some solutions to the above query, it is really important to set the variable $\nabla^2\psi=0$ ref.\cite{I22}. Throughout the present paper, we reduce the subsequent equations associated with the general consideration of Eqs.(1)-(5) while regarding there previous setting.
\\
\\
Following the paper of Piran and al. \cite{I7} expressing the amplitude of incoming and outgoing wave vectors, with the particular metric considered herein with $\nabla^2\psi=0$ ref.\cite{I22}, we obtain:
\begin{equation}
\
A = \frac{{e^{2\psi } }}{\rho}\omega ,_v,
\
 \end{equation}
 and
\begin{equation}
\
B=  \frac{{e^{2\psi } }}{\rho}\omega ,_u.
\
 \end{equation}
We rewrite the  Einstein field equations as a set of four ordinary differential equations coupled according to the characteristics $u$ and $v$ , and we obtain the following expressions:
\begin{equation}
\
A,_u  = \frac{{A+ B}}{{2\rho}},
\
\end{equation}
\begin{equation}
\
B,_v  =  - \frac{{A+ B}}{{2\rho}},
\
\end{equation}
and the function $\gamma$ is determined by
\begin{equation}
\
\gamma ,_\rho  = \frac{\rho}{8}( A ^2  + B ^2 ),
\
\end{equation}
and
\begin{equation}
\
\gamma ,_t  = \frac{\rho}{8}( A^2  - B ^2 ).
\
\end{equation}
We specify that the field equations related to the energy density $\gamma ,_t$ represents the non-gravitational energy density of the wave and
$\gamma ,_\rho$ the gravitational energy density. It is important to note that the introduction of the $v$ and $u$ wave vectors present in Eqs.(6)-(9) plays a fundamental role in clarifying the decomposition of the gravitational wave into explosion and implosion waves during its propagation in spacetime .In this article, we will focus on $B$ and $A$ which play the role of the explosion and implosion wave. We deduce, as mentioned previously, the amplitude of the explosion and implosion radiation $A$ and $B$ wave, respectively. The physical implication is presented in the caption of the figure 1. In this case, we have chosen the radial axis $\rho$, to represent the observables $A$ and $B$. By way of illustration, we plot the energy density $\gamma ,_t $ and $\gamma ,_\rho $ as a function of $\rho$ and $t$ in figures 2 and 3. For figures A1 and A2, we have used the observables $A$ and $B$ to construct a signal that is very close to that of the gravitational wave as observed by the  scientific team LIGO-VIRGO \cite{I21}.
\\
\\
\subsection{Soliton solution analysis}
In this section, we analyze the system of two-soliton solutions cylindrical in the case of the special metric we mentioned, in order to better understand the behavior of the gravitational wave.
\\
\subsubsection{Jordan and Ehlers metric}
Applying the conditions $\nabla^2\psi=0 $ ref. \cite{I22} to the Jordan and Ehlers metric in ref. \cite{I11,I19,I20,I22} governed by Eq.(1) to study the behavior of the gravitational wave evolving according to the $'\times' $ polarization, it is very complex to consider a simplification of Eqs.(1)-(5), for the simple reason that the $\psi $ function has a progressive influence on the propagation of the gravitational wave governed by the $''\times''$ polarization. This equation is actually one particular illustration of the method developed by Belinskii and Zahkarov \cite{I10} when investigating the integration of Einstein equations within the curved space background. As mentioned by these authors in this work, the inclusion of the off-diagonal components such as the ones presented in Eq.(1) with parameter $\omega\neq0$ renders the system more complicated to study in terms of the exact solutions. Therefore, it is very difficult to search for a general N-soliton solutions as in the flat background case \cite{I23}. Using the system of two solutions of cylindrical solitons obtained by Tomizawa and Mishima \cite{I19} in a curved spacetime, whose mathematical foundation is based on (ISM) de Pomeransky's \cite{I18}, we apply the condition $\nabla^2\psi=0$ ref.\cite{I22} to the general solutions to determine the coefficients $e^{2\gamma}$ and $\omega$ of the special metric.
\\
\\
It is clear that the analysis of the cylindrical two-soliton system applied to Einstein's field equations in Xanthopoulos \cite{I24} work has revealed a certain phenomenon and its consequences. In this work, he succeeded in proving that the system of two soliton solutions applied to a rotating cosmic chain in four-dimensional space does not suffer from any pathology concerning the causality violation of spacetime. In this investigation, he has opened up a solid lead regarding a challenge according to which energy should be quantized in the presence of a rotating cosmic string. He presents in this paper that the introduction of the two-solution soliton system makes it possible to obtain regular spacetime in the absence of any singularities, thus leading to a cosmic string surrounded by gravitational waves. According to Xanthopoulos work \cite{I24}, the use of the two-solution cylindrical soliton system as formulated above allows spacetime to be considered as regular, thus facilitating the description of gravitational waves. In the same vein, Francisco \cite{I25} work shows that the fact the spacetime is regular at all points makes it possible to understand that the collision of two massive black holes, each described by four poles of solitons in general relativity, reduces to a system of two solutions of cylindrical solitons, leading to gravitational waves as observed by the scientific team LIGO-Virgo \cite{I21}. It is then important to establish the relationship between the Cartesian coordinates obtained and the cylindrical coordinates. The relationship between the different coordinates can be expressed as follows \cite{I24}:
\begin{equation}
\
x = t - \frac{{\rho ^2 t}}{{2(1 + t^2 )}} + o(\rho ^4 ),
\ \end{equation}
and
\begin{equation}
\ y = 1 + \frac{{\rho ^2 }}{{2(1 + t^2 )}} + o (\rho ^4 ) \,
\end{equation}
for $\rho  \ll \left| t \right|$. In the case where $\rho  \gg \left| t \right|$, we obtain the following expressions:
\begin{equation}
\
y = \rho  + \frac{{1 - t^2 }}{{2\rho }} + o (\rho ^{ - 2} ),
\
\end{equation}
and
\begin{equation}
\
x = \frac{t}{\rho } + \frac{{t(t^2  - 1)}}{{2\rho ^3 }} + o(\rho ^{ - 4} ). \
\end{equation}
In the following subsections, we see the detailed behavior of waves propagating near the limits of spacetime, with a particular focus on waves of explosions and implosions as well as energy densities.
\newline
\subsubsection{Parameter conditions required to study the gravitational wave}
We study the specific behaviour of explosion and implosion waves as well as the notion of energy density. We consider the following conditions: $k = \left| {a_r + ia_i } \right| = \left| a \right|$, $\theta=Arg(a)$ as define in ref.\cite{I19}. Following the same paper were the origin of some parameter $q$ is clearly explained, in the present study we only consider the case $q=1$, because the parameter $q$ can be normalized by a scaling of the coordinates \cite{I12,I19}. We specify that $q$ is a positive constant derived from the two-soliton solution by applying (ISM) Pomeransky \cite{I18} in solving Eqs.(1)-(5). It should be noted that the complex parameter $a$ derived from the two-soliton solution we play the role of gravitational field in the construction and obtaining of the gravitational wave signal, which we discuss progressively. In this investigation, we study the behavior of the gravitational wave in different regions of spacetime under the conditions $t \to \infty$ and $\rho \to \infty$.

\subsubsection{\textbf{Spacelike infinity}}

We study the behaviour of the impulse wave when $\rho\to \infty$ and its gravitational energy density during its propagation and we obtain the following graphs:
\begin{figure}[H]
\begin{center}
\includegraphics[width=9.4cm,height=12cm]{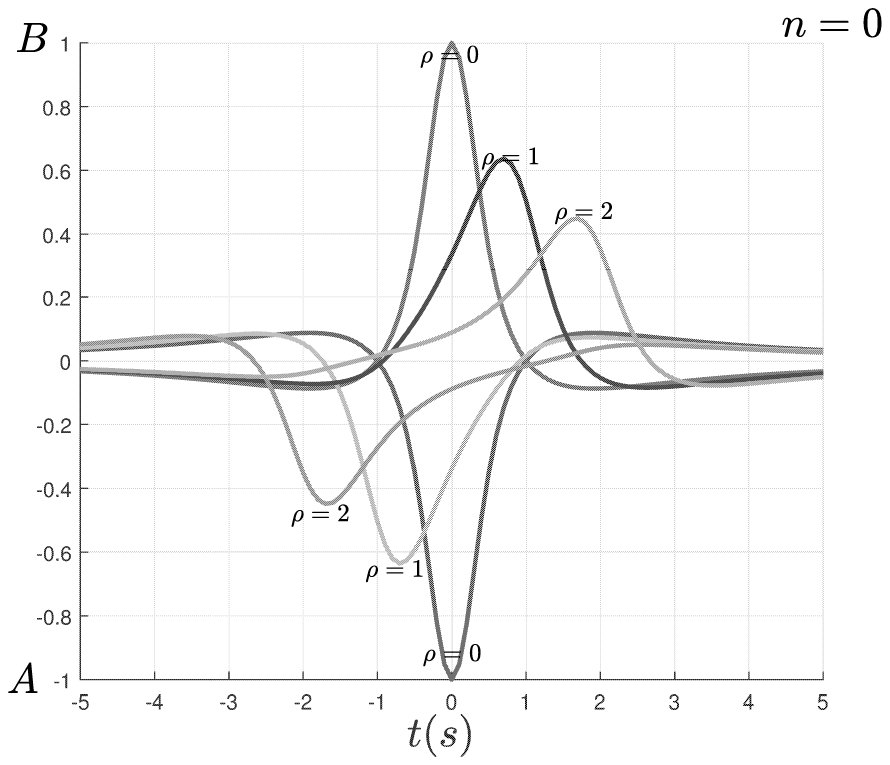}
\caption {We note that at $\rho=0$, the amplitude is maximum.We show the different behaviors of explosives and implosive waves under the following conditions: $(k,\theta,q)=(2,n\pi/4,1)(n=0)$ with $(\rho =0,1,2)$.}
\end{center}
\end{figure}
The result presented in Fig.1 can serve as an ideal test of the presence of different types of gravitational waves in this region as shown in the ref. \cite{I26}. In this analysis, we specify that the propagation of the wave is done in a random way in the absence of different types of noise which allows to obtain a signal close to this study. This observation confirms the hypothesis that the signal of the LIGO-Hanford detector is the inverse of the signal of LIGO-Livingston \cite{I21}. Moreover, it is worth noting that the different amplitudes of the explosion and implosion waves decrease as $\rho\to \infty $, this feature is due to the fact that the waveforms obtained evolve towards an axisymmetric collapse of a rotating black hole, as Stark. \cite{I27} pointed out.

\subsubsection{Discussion of energy density for $t \to \infty$ and $\rho \to \infty$}

The propagation of the impulse wave as an explosion and implosion wave in the regions of space causes an energy density governed by Einstein's field equations. We are interested in each energy density as a function of the region of propagation of the wave. For this, we use the following equations: \begin{equation}
\
\gamma ,_t  = \frac{\rho}{8}( A^2  - B ^2 ),
\
\end{equation}
and
\begin{equation}
\
\gamma ,_\rho  = \frac{\rho}{8}( A ^2  + B ^2 ).
\
\end{equation}

\subsubsection{\textbf{Timelike infinity}}
In this study, we examine the behavior of the gravitational energy density and that of the non-gravitational energy density governed by Eqs.(16)-(17) caused by the wave during propagation in the special case of Jordan and Ehlers spacetime. Using Eqs.(16)-(17), with the parameters previously defined, we obtain the following configurations:
\begin{figure}[H]
\begin{center}
\includegraphics[width=6.3cm,height=12cm]{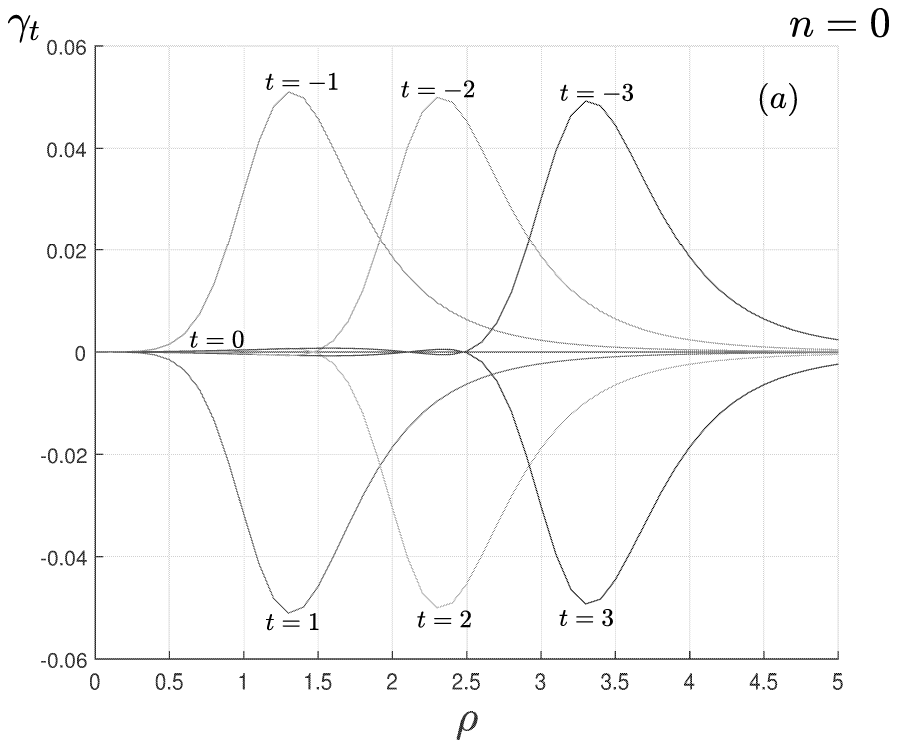}
\includegraphics[width=6.3cm,height=12cm]{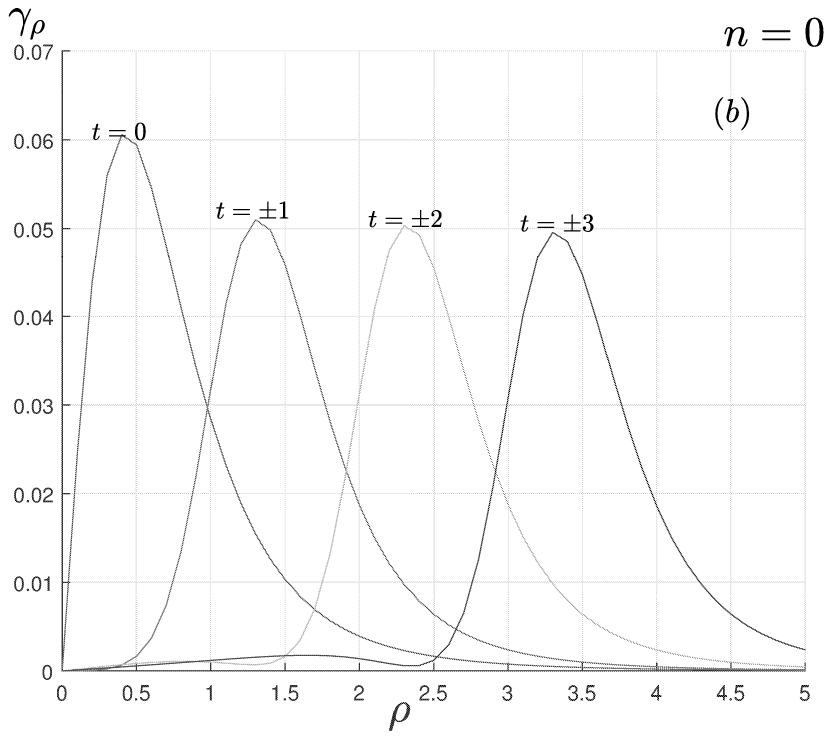}
\caption {$(2a)$: We have the representation of the energy density $\gamma ,_t$ with the following conditions: $(k,\theta,q)= (2,n\pi/4,1)(n=0)$ with $t =  \pm 1, \pm 2, \pm 3$. $(2b)$: We have the representation of the energy density $\gamma ,_\rho$ with the following conditions: $(k,\theta,q)= (2,n\pi/4,1)(n=0)$ with $t =  \pm 1, \pm 2, \pm 3$.}
\end{center}
\end{figure}
The simulation presented in Fig.2 confirms similar results on the case of gravitational waves of Einstein and Rosen widely developed in ref \cite{I8,I28}.
\
\subsubsection{\textbf{Spacelike infinity}}
\
We study the behavior of the gravitational and non-gravitational energy density of the gravitational wave propagating as an explosion and implosion wave governed by Eqs.(16)-(17) in the specific Jordan and Ehlers spacetime \cite{I11}. For this purpose, we use the previously defined parameters to obtain the following configurations:
\begin{figure}[H]
\begin{center}
\includegraphics[width=6.3cm,height=12cm]{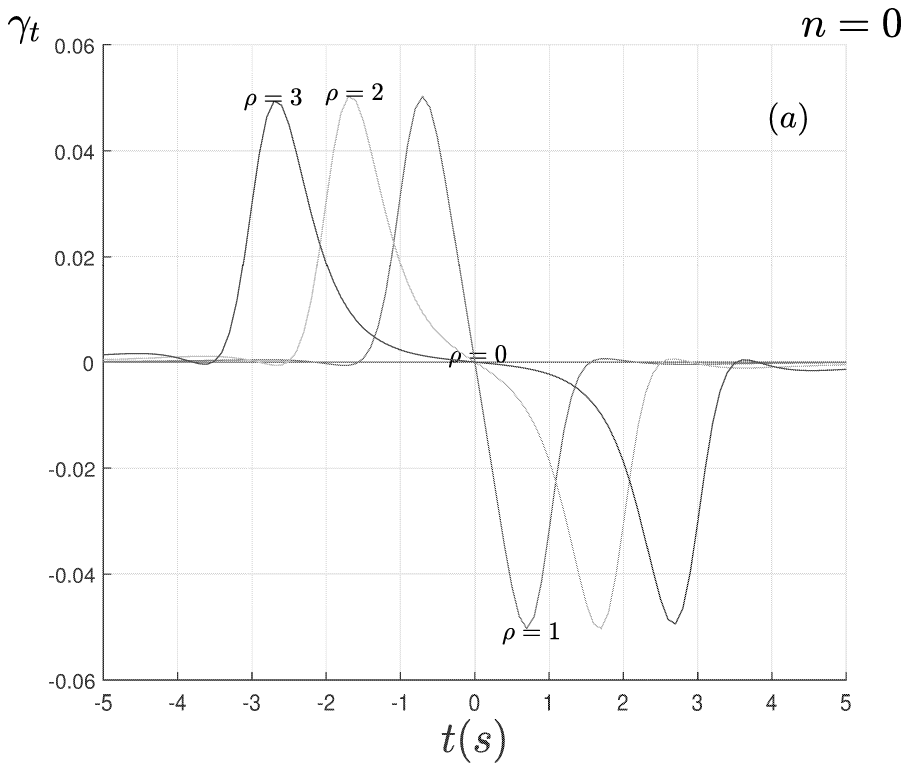}
\includegraphics[width=6.3cm,height=12cm]{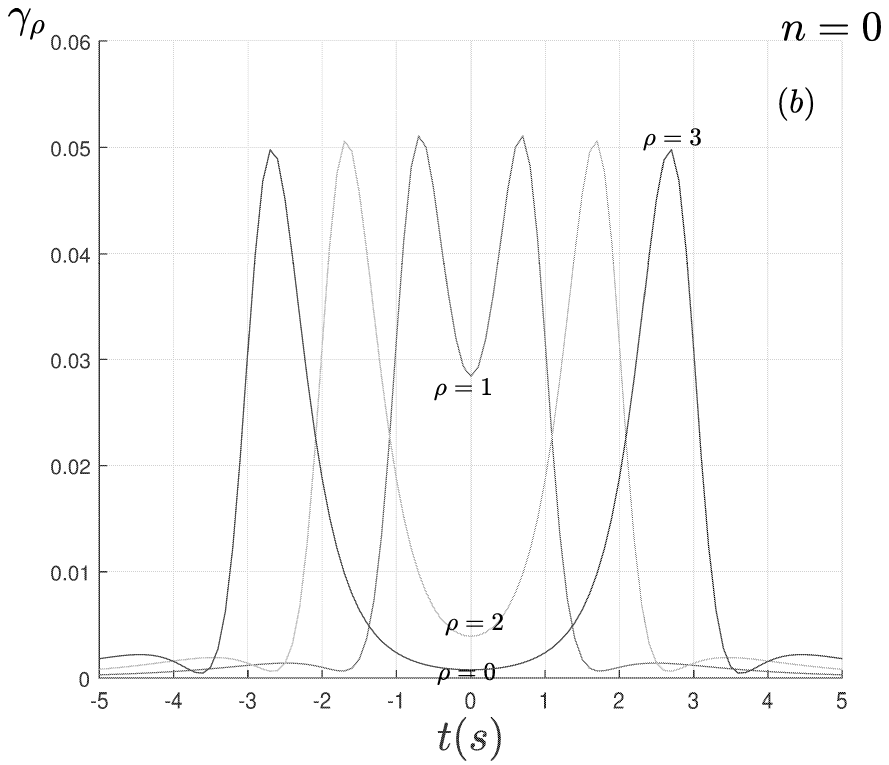}
\caption {$(3a)$: We have the representation of the energy density $\gamma ,_t$ with the following conditions: $(k,\theta,q)= (2,n\pi/4,1)(n=0)$ with $(\rho =0,1,2,3$). $(3b)$: We have the representation of the energy density $\gamma ,_\rho$ with the following conditions: $(k,\theta,q)= (2,n\pi/4,1)(n=0)$ with $(\rho =0,1,2,3$).}
\end{center}
\end{figure}
Fig.3 confirms that in the ideal test area, the gravitational wave carries an energy density as contained in the literature \cite{I7,I8,I28,I29,I30,I31}.
It is important to point out that varying the coordinates $\rho$ and $t $ in Fig.2 and Fig.3  gives a different interpretation of the gravitational wave's behavior in relation to different energy densities. This leads us to note that the non-gravitational energy density shown in Figs.(2a) and Fig.(3a) alternates between positive and negative values, while the gravitational energy density shown in Figs. (2b) and Fig.(3b) remains positive. It should be emphasized that the different densities studied exist when $\rho\rightarrow0$ and $t\rightarrow0$, even though they are extremely low in Figs.(2a), Figs.(3a) and Figs.(3b). In studying the various phenomena highlighted in Fig.1, Fig.2, and Fig.3, we were obliged to choose the value $n=0$ to obtain the smallest angle of rotation $\theta=0$ of the black hole, since the higher $\theta$ is, the more the amplitudes of the explosion and implosion waves are divided. Moreover, introducing the modulus $k$ into the analysis allows us to quickly simplify the cumbersome expressions contained in two-soliton solutions.
\newline
\newline
Considering the different results obtained in the discussion of the energy density of the impulse wave during its evolution in the considered spatial regions, we obtain the unique conditions on the energy ($\gamma ,_t\neq0 $ and $\gamma ,_\rho\neq0 $) for $a\neq0 $. For $a=0$, the different energy densities relative to the propagation of the impulse wave take the form of transcendental cylindrical waves \cite{I32} $\gamma ,_{t\rho }=0$.
\newline
\newline
\subsection{CONCLUSION}
In this paper, we studied the cylindrical gravitational pulse wave particular case  in   Jordan and Ehlers spacetime \cite{I11}. In our investigations, we used the method of Piran and al. \cite{I7} to solve the Einstein field equations. The method of Piran and al. \cite{I7}  describe two types of waves, namely linear and nonlinear waves. We mentioned that the linear waves resulting from the numerical solutions, represent the waves of Einstein and Rosen \cite{I3} which in the work of Weber and Wheeler \cite{I4} are present in the form of explosions and implosions waves, respectively. In this dynamics, we constructed an impulse wave in specific  the spacetime of Jordan and Ehlers \cite{I11} in the form of explosion and implosion wave which we assimilate to the gravitational soliton thanks to the ISM of Pomeransky \cite{I18}. In the construction of the impulse wave, we showed that it presents similar characteristics to the one studied ref \cite{I4,I6,I7,I8} concerning the different densities of energies, as well as the shapes of the wave in the spacetime.  The use of the method of Piran and al. \cite{I7}, allows us to showed that the nonlinear cylindrical gravitational impulse wave, which speed is close to that of light, moves in the form of two radiations, namely: the explosive radiation and the implosive radiation,respectively. In this paper, we insist that $B$ and $A$ represent the explosive and implosive wave because this respects the Weber and Wheeler \cite{I4} analogy.  We realized that, when the complex-valued $a=0$, we have the following different characteristics of the nonlinear gravitational impulse wave: $B=A=0$, $\gamma ,{t\rho }=0$ and with $\nu ^\rho$ not defined. Using $\gamma ,_{t\rho }=0$, this condition found by Chandrasekhar \cite{I32} shows that the gravitational soliton behaves like transcendental cylindrical waves with conserved energy. Considering $a\ne 0$, we obtained the wave characteristics : $A\ne 0$, $B \ne 0$, $\gamma ,_{t\rho }\ne 0$ and $\nu ^\rho \ne 0$. By respecting the condition $a\ne 0$, we obtain the Halilsoy criterion \cite{I8} $\gamma ,_{t\rho }\ne 0$ on the gravitational waves of Einstein and Rosen. We note that the gravitational wave contained, in the nonlinear spacetime admit similarities with the works contained in the literature \cite{I4,I7,I8,I28,I29,I30,I31}. It is also important to note that in this article the combination of the method of Piran et al. \cite{I7} and (ISM) of Pomeansky \cite{I18} opens a possibility of detection of gravitational waves as revealed in the work of Francisco \cite{I25}.
\\
\begin{appendices}
 \section{ Gravitational Waves}\label{secA1}
In this section, we examine the possibility that the gravitational soliton we have assimilated to the gravitational wave in the form of an explosion and implosion wave could be observed by the LIGO-Virgo-KAGRA detectors \cite{I31}. Before examining this possibility, it is important to return to the condition we set, namely $\nabla^2\psi=0$ applied to Eqs.(1)-(5) in order to examine the behavior of the gravitational wave evolving according to the $''\times'' $ polarization. Clearly, the $\nabla^2\psi=0$ condition applied to Eq.(1) translates the behavior of a cylindrical gravitational wave evolving in two different dynamics, namely: the stationary regime and the non-stationary regime, as pointed out by Stachel \cite{I33}. In the stationary regime, we obtained Chandrasekhar's \cite{I32} transcendental cylindrical wave with the properties ($\gamma ,_{t\rho }=0$, $a=0$ and $\psi=\omega=0$) which allows us to simplify Eq.(1) into the following form:
\begin{equation}
ds^2  = e^{2\gamma} (d\rho ^2  - dt^2 ) + \rho ^2  d\phi ^2  + dz^2.
\end{equation}
This equation was examined by Weber and Wheeler \cite{I4} where they demonstrated that the gravitational energy radiated by Eq.(A1) is positive and constant. However, in the non-stationary regime, the gravitational wave evolves as an Einstein and Rosen wave \cite{I8}. Clearly, if the theory of general relativity is true, as Throne \cite{I34} suggests, then the various detectors \cite{I21,I26} must be able to observe the different signals from the $''+''$ and $''\times''$ polarizations.
\\
\subsection{Detector of gravitational waves (LIGO-Virgo-KAGRA signals)}
In this subsection, we focus on building and obtaining gravitational waves propagating along the $''\times' $ polarization. For this, we rely on the construction of gravitational waves propagating along the $''+' $ \cite{I31} polarization. In this investigation, we need to establish the rules and conditions required to obtain the various signals mentioned in ref.\cite{I21,I26}. Although we use analytical solutions to solve the problem of constructing gravitational waves to the detriment of numerical solutions \cite{I27}, in this case the following conditions would have to be met: \textbf{1}) the regularity of the gravitational field in the detection zone, \textbf{2}) the ideal rotation angle ($\theta=0$) of the black holes and \textbf{3}) the perturbations linked to the gravitational field. Beyond the conditions imposed, in the specific case of gravitational wave propagation governing the $''\times''$ polarization, it is important that the gravitational field tends to be static so that the $''\times''$ mode is pure, verifying a Stachel \cite{I33} condition. Using this procedure, the expressions relating to Eqs.(6)-(7) characterizing the explosion and implosion waves are represented and reduced to the following form:
\begin{equation}
\
A \simeq \frac{\omega}\rho ,_v,
\
 \end{equation}
 and
\begin{equation}
\
B\simeq \frac{\omega}\rho ,_u.
\
\end{equation}
By taking into account Eqs.(A2)-(A3), the modulus $k\rightarrow0$  quickly simplifies the heavy expressions contained in the two-solitons solution, facilitating analysis of the behavior of the gravitational wave evolving on pure polarization $(''\times'')$. In view of the above arguments, it is important to stress that the accuracy of the construction and obtaining of the gravitational wave in the various detectors in this study depends on the complex parameter $a $ and the notion of noise \cite{I21,I26,I31}. Using Eqs.(A2)-(A3) representing the explosion and implosion waves and the perturbation of the gravitational field due to white Gaussian noise, we obtain the signals from the various detectors shown in the following figures:
\begin{figure}[H]
\begin{center}
\includegraphics[width=16.4cm,height=18cm]{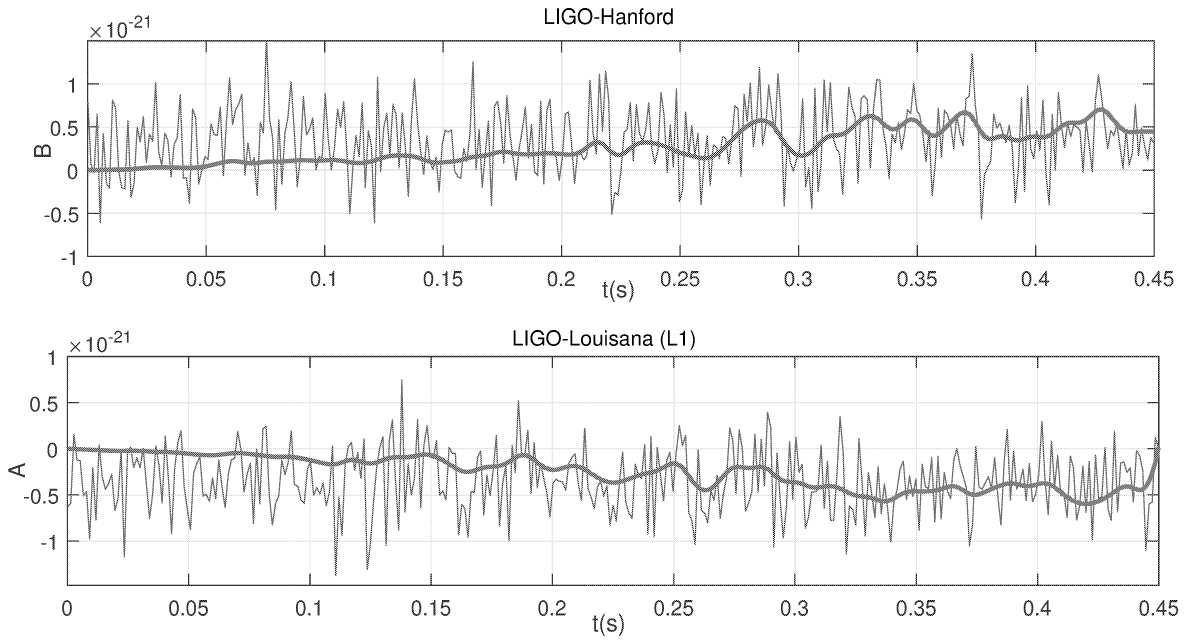}
 \caption {Profile $B$ presenting an explosion signal wave with the introduction of a Gaussian noise of the order of $1.01dB$. We use the following parameters: for $(k,\theta,q)= (0.02,0,1)$ with
$(\rho =3001111)$. This feature is actually close to the signals observed by the LIGO-Hanford \cite{I21}.}
\caption {Profile $A$ presenting an implosion signal wave with the introduction of a Gaussian noise of the order of $1.01dB$. We use the following parameters: for $(k,\theta,q)= (0.02,0,1)$ with
$(\rho =3001111)$. This feature is actually close to the signals observed by the LIGO-Livingston \cite{I21}.}
\end{center}
\end{figure}
Looking at figs A1 and A2, we can see the presence of two signals, notably the blue signal, which characterizes the random passage of the gravitational wave through the LIGO-Virgo-KAGRA detector, whose waveform is governed by Gaussian noise due to the presence of gravitons during the propagation of the gravitational wave in the interferometer \cite{I35}. The red curve is obtained after reconstruction of the gravitational wave signal. This investigation confirms that the collision of two massive black holes in general relativity theory leading to the detection of gravitational waves by the various interferometers corresponds to two types of mode, namely the $''+'''$ mode and the $''\times'''$ mode, as confirmed by ref.\cite{I34,I36}. This shows that each mode is indeed capable of producing a gravitational wave signal ref. \cite{I31} in the presence of Gaussian noise as obtained by the LIGO-VIRGO scientific team  \cite{I21,I26}. Taking into account the remarks of Abramovici et al. \cite{I36}, we can see that it is impossible to obtain this type of gravitational wave if spacetime is flat. We note that the analytical approach used in this investigation is similar to other numerical methods exploited in the construction and observation of gravitational waves, with the similar waveforms developed in Fig. 1 \cite{I26,I27}. In this dynamic, we can see that the data processed throughout this work correspond to the behavior of the collapse of a rotating black hole, as confirmed in Fig. 1 \cite{I27}. What's more, this experiment confirms that gravitational wave signals from black holes take only a few seconds in detectors \cite{I26} before disappearing completely. We note that the analytical approach offers a facility over Bayesian parameter estimation \cite{I26} methods for the following reasons: \textbf{1}) it requires fewer data analysis parameters, \textbf{2}) it requires the solution of the Einstein field equations to be exact and \textbf{3}) it requires the gravitational field to be regular in all detection zones. One of the difficulties of this method lies in the fact that the perfection of the signals observed in the different detectors depends solely on the random test of the gravitational field, which is extremely complex to achieve. On the other hand, Bayesian parameter estimation methods \cite{I21,I26} have the advantage that, when applied to exact wave models, they process the data accurately, automatically adjusting and subtracting the data while establishing a well-defined confidence region. One of the difficulties of this method is that it only applies to exact wave models. Beyond this condition, it requires the use of several parameters in the analysis and processing of data, which in turn necessitates the use of multiple algorithms to establish the correlation in the confirmation of the processed data \cite{I21,I26}. This result is in pretty agreement with the recent works of the scientific team of Ligo-Virgo \cite{I21,I26}. Such a result allows us to confirm that (ISM) Pomeransky’s \cite{I18} conveniently and completely describes the gravitational pulse wave compared to other methods. In order to generate the previous characteristics, we introduced the Gaussian noise expressions into the field equations according to the usual procedure \cite{I31}. As far we are concerned, how to more precisely adjust the perturbations of the gravitational field from the scale to that of the detector LIGO-Virgo-KAGRA remains a major current challenge.
\\
\\
\end{appendices}
\\
\textbf{Declaration: The authors did not receive support from any organization for the submitted work. No funding was received to assist with the preparation of this manuscript.}
\\
\\

\end{document}